\documentclass[%
reprint,
 amsmath,amssymb,
 aps,
]{revtex4-2}

\usepackage{graphicx}
\usepackage{dcolumn}
\usepackage{bm}
\usepackage{hyperref}
\usepackage{color}

\usepackage[T1]{fontenc}

\DeclareRobustCommand{\VAN}[3]{#2}
\let\VANthebibliography\thebibliography
\def\thebibliography{\DeclareRobustCommand{\VAN}[3]{##3}\VANthebibliography}

\usepackage{graphicx}	
\usepackage{amsmath}	

\begin{document}



\title{Impact of the nuclear equation of state on the explodability of massive stars}

\author{Jade Powell\thanks{E-mail: dr.jade.powell@gmail.com}}
\affiliation{
Centre for Astrophysics and Supercomputing, Swinburne University of Technology, Hawthorn, VIC 3122, Australia
}
\author{Bernhard M\"uller\thanks{E-mail: bernhard.mueller@monash.edu}}
\affiliation{%
School of Physics and Astronomy,
10 College Walk,
 Monash University,
 Clayton, VIC 3800,
 Australia
}



\label{firstpage}

\begin{abstract}
In recent years, astrophysical observations have placed tight constraints on key properties of the nuclear equation of state (EoS). Using 93 two-dimensional simulations for three different EoS compatible with the current tight constraints, we show that the EoS remains a major uncertainty for the outcome of core-collapse supernovae. Whereas explosions are obtained in most cases for the SFHo and SFHx EoS, for the CMF EoS, which includes a crossover from nucleonic matter to a quark phase, explosions occur only for 2 out of 15 progenitors. Less favourable conditions for neutrino-driven explosions arise for the CMF EoS due to lower neutrino luminosities and mean energies and slightly weaker contraction of the warm proto-neutron star. Our results suggest that the explodability of massive stars cannot yet be predicted based on first principles without better knowledge of the nuclear EoS. Conversely, observational constraints on stellar explodability may help further constrain the EoS.
\end{abstract}

\maketitle


\begin{table*}
\centering
\begin{tabular}{ c c c c c c c c c c}
\hline\hline
 \bf{} & \bf{Duration} & \bf{Energy} & \bf{Revival} & \bf{Duration} & \bf{Energy} & \bf{Revival} & \bf{Duration} & \bf{Energy} & \bf{Revival}  \\  
 \bf{Model} & \bf{CMF (s)} & \bf{CMF (0.1\,B)} & \bf{CMF (s)} & \bf{SFHo (s)} & \bf{SFHo (0.1\,B)} & \bf{SFHo (s)} & \bf{SFHx (s)} & \bf{SFHx (0.1\,B)} & \bf{SFHx (s)} \\
  \hline
 s9.71  & 4.95 & 0.46 (0.40) & 0.253 & 2.79 & 0.82 (0.82) & 0.165 & 4.41 & 0.64 (0.53) & 0.168 \\ \hline
 s10.13 & 4.99 & No exp.  & N/A   & 2.78 & 1.01 (0.91) & 0.129 & 4.40 & 0.81 (0.67) & 0.129 \\ \hline
 s10.6  & 5.26 & No exp.  & N/A   & 2.43 & 0.68 (0.62) & 0.157 & 4.39 & 0.45 (0.29) & 0.207 \\ \hline
 s11.5  & 5.01 & No exp.  & N/A   & 2.73 & 0.16 (0.16) & 0.404 & 2.82 & 0.26 (0.26) & 0.361 \\ \hline
 s12.5  & 5.33 & No exp.  & N/A   & 2.71 & 1.19 (0.98) & 0.274 & 4.28 & 1.48 (1.13) & 0.247 \\ \hline
 s13.11 & 5.41 & No exp.  & N/A   & 2.83 & 2.06 (1.73) & 0.327 & 3.83 & 5.70 (3.28) & 0.375 \\ \hline
 s14    & 4.97 & No exp.  & N/A   & 2.78 & 2.46 (2.37) & 0.270 & 4.61 & No exp.  & N/A \\ 
 s14a & 1.59 & No exp. & N/A & 1.95 & 2.70 (2.65) & 0.251 & 1.93 & 1.29 (1.29) & 0.229 \\
 s14b & 1.60 & No exp. & N/A & 1.97 & 3.94 (3.90) & 0.224 & 1.97 & 2.50 (2.46) & 0.265 \\
 s14c & 1.15 & No exp. & N/A & 1.91 & 2.07 (2.07) & 0.303 & 1.93 & 3.57 (3.57) & 0.270 \\
 s14d & 1.20 & No exp. & N/A & 1.92 & 2.97 (2.97) & 0.275 & 1.95 & 3.66 (3.57) & 0.314 \\ \hline
 s15    & 4.74 & No exp.  & N/A   & 3.08 & No exp.  & N/A   & 4.91 & No exp.  & N/A \\ 
 s15a & 1.24 & No exp. & N/A & 1.95 & 5.33 (5.19) & 0.284 & 1.95 & No exp. & N/A \\
 s15b & 1.18 & No exp. & N/A & 1.96 & No exp.     & N/A   & 1.98 & No exp. & N/A \\
 s15c & 1.56 & No exp. & N/A & 1.92 & No exp.     & N/A   & 1.93 & 4.70 (4.68) & 0.236 \\
 s15d & 1.20 & No exp. & N/A & 1.93 & 3.61 (3.55) & 1.117 & 1.95 & No exp. & N/A \\ \hline
 s16.22 & 4.54 & No exp.  & N/A   & 2.72 & 0.41 (0.27) & 0.263 & 4.20 & 1.62 (0.66) & 0.296 \\ \hline
 s18    & 2.04 & No exp.  & N/A   & 2.91 & No exp.  & N/A   & 4.15 & 6.57 (3.31) & 0.324 \\ 
 s18a & 1.58 & No exp. & N/A & 1.96 & 5.56 (5.64)   & 0.270 & 1.95 & 4.42 (4.42) & 0.233 \\
 s18b & 1.56 & No exp. & N/A & 1.96 & 4.58 (4.55)   & 0.218 & 1.98 & No exp. & N/A \\
 s18c & 1.54 & No exp. & N/A & 1.95 & 17.11 (16.99) & 0.263 & 1.97 & 3.38 (3.32) & 0.315 \\
 s18d & 1.22 & No exp. & N/A & 1.94 & 13.41 (12.98)  & 0.310 & 1.94 & 2.07 (2.10) & 0.423 \\ \hline
 s19.5  & 4.59 & 9.28 (2.51) & 0.291  & 2.64  & 5.33 (3.40) & 0.256 & 3.96 & 1.45 (0.69) & 0.287 \\ 
 s19.5a & 1.23 & No exp. & N/A & 1.92 & 3.79 (3.79) & 0.333 & 1.92 & 1.08 (1.07) & 0.386 \\
 s19.5b & 1.24 & No exp. & N/A & 1.92 & 4.05 (4.05) & 0.281 & 1.91 & 0.74 (0.74) & 0.240 \\
 s19.5c & 1.21 & No exp. & N/A & 1.92 & 2.32 (2.32) & 0.326 & 1.91 & 7.08 (7.08) & 0.282 \\
 s19.5d & 0.82 & No exp. & N/A & 1.97 & 4.06 (4.06) & 0.373 & 1.92 & 3.36 (3.28) & 0.302 \\ \hline
 s21.91 & 1.67 & No exp.  & N/A   & 2.65 & 4.26 (3.07) & 0.243 & 3.88 & 13.0 (4.38) & 0.287 \\ \hline
 s24    & 3.19 & No exp.  & N/A   & 2.44 & 14.8 (9.84) & 0.279 & 3.93 & 3.68 (2.09) & 0.456 \\ \hline
 s29.59 & 1.35 & No exp.  & N/A   & 2.68 & 11.7 (11.2) & 0.308 & 4.05 & 18.5 (17.8) & 0.312 \\ \hline
 s36.61 & 0.70 & No exp.  & N/A   & 1.91 & 33.7 (33.7) & 0.247 & 3.64 & 15.9 (16.7) & 0.222\\
 \hline\hline
\end{tabular}
\caption{The first column is the progenitor model, labelled as ``s'' plus the ZAMS mass. For each EoS, we show the duration after bounce when the simulation stops, the diagnostic explosion energy at the end of the simulation and in brackets at 1.91\,s in units of $0.1\,B=10^{50}\,\mathrm{erg}$, and the time of shock revival. We define the shock revival time as the time when the average shock radius reaches 250\,km. Only two of the models with the CMF EoS undergo shock revival. 
We vary only the progenitor mass, EoS, and the random seed for the initial perturbations in each model.}
\label{tab:energies}
\end{table*}

\section{Introduction}
Core-collapse supernovae (CCSNe) are the explosive deaths of stars larger than $\sim 8\,\mathrm{M}_{\odot}$. Massive stars undergo several stages of thermonuclear fusion until they form an iron core, which collapses into a proto-neutron star (PNS) when it approaches the effective Chandrasekhar mass. The rebound of the collapsing core launches a shock wave, which initially stalls before it is revived and expels the stellar envelope. For typical CCSNe, shock revival is thought to occur through the reabsorption of energy from  neutrinos emitted from the PNS \citep{janka_17,mueller_20,mezzacappa_20}.
For more extreme CCSNe, the energy may come from the rotation and magnetic fields \citep{burrows_07,mosta_14,reichert_23,muller_24}. 

In recent years, multi-dimensional simulations with neutrino transport have become efficient enough to explore the systematics of CCSN explosion and compact remnant properties \citep{nakamura_15, mueller_19, burrows_20},
complementing simpler, phenomenological or lower-dimensional models \citep{ugliano_12, sukhbold_16, mueller_16, couch_20, ghosh_22}
that can still scan the parameter space of progenitors more thoroughly. Many of the qualitative findings from these studies are likely robust consequences of CCSN physics. It is imperative, however, to determine the sensitivity of the
qualitative and quantitative outcomes of CCSN explosions to the numerous ingredients that enter this complex multi-physics problem.

One of the key ingredients is the nuclear equation of state (EoS).
The EoS is known to affect the contraction of the PNS, and hence
neutrino emission, neutrino heating and the conditions for shock
revival \citep{janka_01, mueller_15}.
Its impact on the PNS structure may possibly be
probed with gravitational waves in the future \citep{bizouard_21, powell_22, mitra_24}.
The EoS also determines the maximum mass of the PNS before it collapses to a black hole. Understanding which stars undergo shock revival, which stars form neutron stars, and which stars will form black holes is one of the major unknowns of CCSN theory, with implications for many areas of astrophysics.
The answers to these questions will be essential for understanding the neutron stars and black holes observed through gravitational-wave \citep{gwtc-4} and electromagnetic observations \citep{ ozel_16, you_25}.
The main goal of this work is to determine if 
the EoS
remains a major factor of uncertainty for theoretical predictions of the ``explodability'' of massive stars.

Studies of EoS effects on CCSN explosions have a long history. Already more than a decade ago, two-dimension studies \citep{janka_12,suwa_13} had demonstrated that EoS that result in faster PNS contraction provide more favourable conditions for explosion. \citet{yasin_20} identified the nucleon effective mass as the major factor that determines the contraction of warm PNSs, which therefore affects explodability. Numerous other groups have also investigated the impact of the EoS on black hole formation in very massive stars \citep{pan_18, powell_21, meskhi_22, Anderson_25}, mass loss in failed CCSNe \citep{ivanov_21}, and the gravitational-wave emission \citep{marek_09, richers_17, sotani_19, Anderson_21, jakobus_23, murphy_24}.

However, a critical impact of the EoS on explodability has so far been demonstrated for EoS with rather substantial differences in key nuclear physics and neutron star properties. Classical results \citep{janka_12,suwa_13,yasin_20}
were obtained for the EoS of \citet{lattimer_91}, sometimes even with a low bulk incompressibility modulus of $K=180\,\mathrm{MeV}$, and that of \citet{shen_98} with an unrealistic symmetry energy.
Recent constraints from theory
\citep{drischler_17}, nuclear experiments \citep{lattimer_13}, and observations of neutron stars in binary systems \citep{raithel_19}
now place rather tighter constraints on the EoS already, and have 
ruled out the  $K=220\,\mathrm{MeV}$ EoS of  \citet{lattimer_91}, which
had enjoyed popularity in CCSN simulations.

Various EoS consistent with current constraints are now available for CCSN simulations, such as SFHo and SFHx EoS from \citet{Steiner_13}, and more recently the CMF EoS from \citet{Motornenko_20} and the DD2 EoS from \citet{fischer_18}. It is therefore critical to reinvestigate whether uncertainties in the  EoS still substantially impact CCSN outcomes despite the much tighter current experimental, observational and theoretical constraints.

In this work, to demonstrate that the EoS still has a significant impact on CCSN outcomes, we consider the impact of the EoS on the PNS evolution and explosion dynamics using 93 axisymmetric simulations with the neutrino hydrodynamics code CoCoNuT-FMT \citep{mueller_15}. We use three different EoS, SFHo, SFHx and CMF, which fall within the limits given by observational constraints. We use a range of different mass progenitor stars from $9.71\,\mathrm{M}_{\odot}$ to $36.61\,\mathrm{M}_{\odot}$.

\section{Progenitor models and simulation methodology}
\label{sec:setup}
We perform a suite of simulations using three different EoS at high densities that fit within the current experimental and observational constraints. These are the CMF EoS from \citet{Motornenko_20}, which includes a crossover to deconfined quark matter at high densities, and the SFHo and SFHx EoS from \citet{Steiner_13}, which only consider nucleons and have different symmetry energies.
The CMF EoS is very stiff up to $5-6\times10^{14}\,\mathrm{g}\,\mathrm{cm}^{-3}$, which allows it to comply with neutron star mass and radius constraints. It then significantly softens above that density (cp. Figure 8 in \cite{jakobus_22}). This means that for all but very massive warm PNS, the contraction is slowed down compared to the SFHo or SFHx EoS. 
The maximum neutron star mass is $2.10\,\mathrm{M}_{\odot}$ for CMF, $2.059\,\mathrm{M}_{\odot}$ for SFHo, and $2.13\,\mathrm{M}_{\odot}$ for SFHx. 
At low densities, we use an EoS for photons, electrons, positrons and an ideal gas of nuclei with a flashing treatment for nuclear reactions \citep{rampp_02}. 
We use 15 different single-star, solar-metallicity progenitor models from \citet{mueller_16,mueller_25}
obtained with the stellar
evolution code \textsc{Kepler} \citep{weaver_78, heger_10}.
The models have zero age main sequence (ZAMS) masses between $9.71\,\mathrm{M}_{\odot}$ and $36.61\,\mathrm{M}_{\odot}$, as listed in Table~\ref{tab:energies} and denoted by the model labels (s9.71 to s36.61). 
We perform our simulations using the neutrino hydrodynamics code \textsc{CoCoNuT-FMT} \citep{mueller_15,mueller_19}. \textsc{CoCoNuT-FMT} combines a general relativistic finite-volume solver with higher-order reconstruction for the equations of hydrodynamics \citep{mueller_10} with the fast multi-group transport (FMT) method for treating the neutrinos \citep{mueller_15}.   
With few exceptions, the models were simulated for several seconds 
(Table~\ref{tab:energies})
to confidently decide whether shock revival occurs or not.


\begin{figure*}
\includegraphics[width=\textwidth]{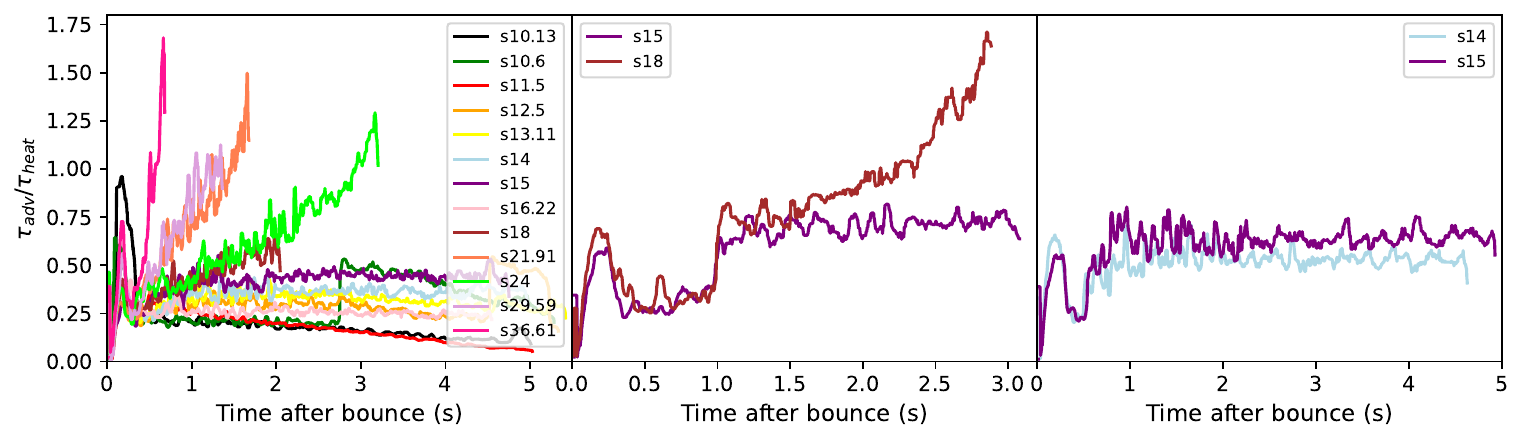}
\caption{The timescale criterion $\tau_{adv} / \tau_{heat}$ for the non-exploding CMF models (left), SFHo models (middle), and SFHx models (right). 
Rapid increase above 1 usually indicates neutrino-driven runaway shock expansion. However, the models with values above 1 formed black holes before shock revival could occur.
}
\label{fig:ratio}
\end{figure*}

\section{Results}
\label{sec:dynamics}
Key outcomes of the simulations, i.e., the time of shock revival (if applicable) and the explosion energy are shown in
Table~\ref{tab:energies}, and differ markedly between the three different EoS. We calculate the diagnostic explosion energy using the method of \cite{mueller_17}.

All of our simulations include initial random seed perturbations in radial velocity. Perturbations are required to break spherical symmetry and drive an explosion. 
To determine the impact of stochasticity, we include an extra four simulations each for progenitors s14, s15, s18 and s19.5 where only the random perturbations were changed. We changed only the random seed, and not the magnitude of the perturbations.
Shock revival is seen in the majority of SFHo models with the exception of 3 of the s15 progenitors and one of the s18. 
The SFHx series shows very similar results, with some failed explosions for models s15 and one for s18, but with the addition of one failed explosion for s14.
In stark contrast, all but two of the CMF models fail to explode. The only exception with successful explosions are the  $9.71\, \mathrm{M}_{\odot}$ progenitor with a very small iron-silicon core mass \citep{mueller_25}, and one of the $19.5\,\mathrm{M}_{\odot}$ progenitor models.  
The one exploding s19.5 CMF model may be just a random artifact, as there was no shock revival observed in the other s19.5 CMF models. 
In most cases, shock revival is associated with the infall of the Si/O shell interface, which typically coincides with the mass shell where the entropy exceeds $4k_B$/nucleon \citep[cf.\ ][]{janka_12,ertl_16,mueller_20}. Some progenitors, such as s13.11, do not have clearly defined shell interfaces, though, but the time of explosion still corresponds to this entropy threshold. For some of the more massive progenitors like s24, and also some of the s18 models, the explosion is delayed relative to the accretion of the shell interface. The stochasticity has the largest impact on models that are closest to the threshold between explosion and non-explosion, for example model s15, and will impact the final outcome of the supernova remnant. For models with more robust explosions, such as model s19.5, we find stochastic variations in the shock revival time of up to 146\,ms. 

To ascertain that the different outcomes between the three EoS are the result of robust differences in the conditions for shock revival, Figure~\ref{fig:ratio} shows the ratio of the advection time scale $\tau_\mathrm{adv}$ through the gain (neutrino heating) region, and the heating time scale $\tau_\mathrm{heat}$ for negating the binding energy of the material in the gain region for the current neutrino heating rate. 
This time-scale ratio quantifies how close the heating conditions are to self-sustained neutrino-driven shock expansion, which is expected around $\tau_\mathrm{adv}/\tau_\mathrm{heat}\approx 1$ \citep{buras_06}. 

In the vast majority of the non-exploding CMF models, the ratio $\tau_\mathrm{adv}/\tau_\mathrm{heat}$ indeed remains well below unity and does not exhibit an appreciable growing trend that might still indicate a late explosion.
In models s21.91, s24, s29.59 and s36.61 the ratio
is growing steadily by the end of the simulation and even exceeds unity shortly before the end. A later explosion can be excluded, however, as these models collapse to black holes at the end of the simulation.
One should note that
$\tau_\mathrm{adv}/\tau_\mathrm{heat} \gtrsim 1$ become a less reliable indicator for shock revival as relativistic effects at the base of the gain region become strong close to black hole formation.

The time-scale criterion also indicates that models s14 and s15 are robust non-explosions for the SFHx EoS, with values $\tau_\mathrm{adv}/\tau_\mathrm{heat}\lesssim 0.7$ and no increasing trend over several seconds. Model s14 is also a robust failure for the SFHo EoS. Model s18 reaches high values of $\tau_\mathrm{adv}/\tau_\mathrm{heat}$ 
with an increasing trend at the end of the simulation and appears close to an explosion. However, it forms a black hole before shock expansion sets in.

\begin{figure*}
\includegraphics[width=\textwidth]{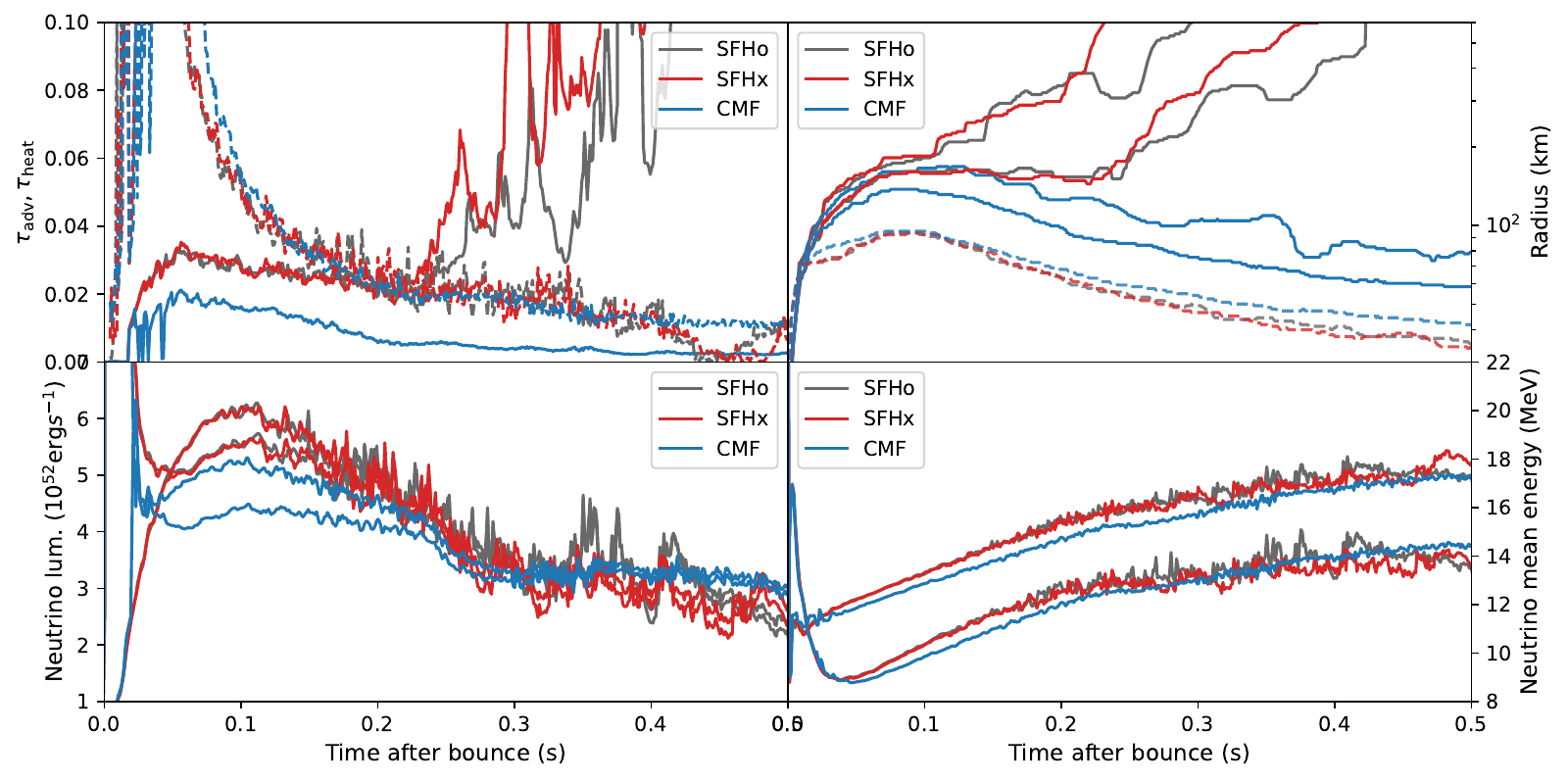}
\caption{A comparison of the underlying differences between the different EoS for the s12.5 model. Top left shows the advective timescale (solid line) and heating timescale (dashed line).
Top right is the minimum and maximum shock and gain radius. Bottom left is shows the neutrino luminosity. Bottom right is the neutrino mean energy.  }
\label{fig:ratio2}
\end{figure*}

The evolution of the time-scale ratio thus demonstrates significant differences in the heating conditions and underscores the robust influence of EoS differences, in particular between CMF vis \`a vis SFHo and SFHx, on CCSN outcomes. In Figure~\ref{fig:ratio2}, we further illuminate the physical causes underlying these differences using the s12.5 models 
as a representative example for low to moderate progenitor masses. The key difference for the CMF model is the shorter advection time scale, whereas the heating time scale is very similar to SFHo and SFHx until the SFHo and SFHx models explode.

\begin{figure*}
    \centering
    \includegraphics[width=\linewidth]{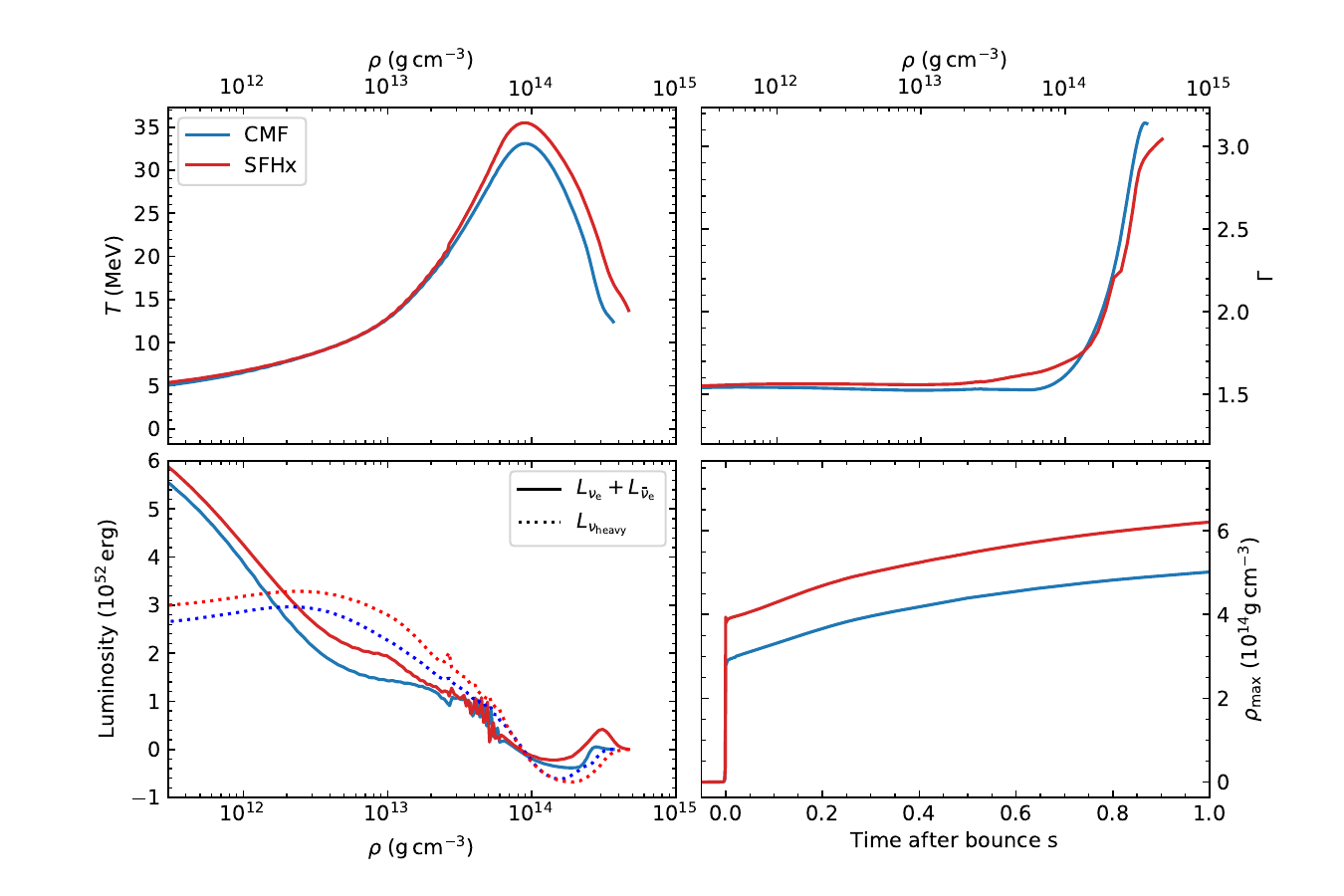}
    \caption{Illustration of differences in the PNS
    structure between the CMF and SFHx EoS for model s12.5.
    The panels compare the temperature $T$ (top left), adiabatic index $\Gamma$ (top right), neutrino luminosity (bottom left) at a post-bounce time of $200\,\mathrm{ms}$. We show the combined luminosity of $L_{\nu_\mathrm{e}}+L_{\bar{\nu}_\mathrm{e}}$ of electron-flavor neutrinos and the luminosity of heavy-flavour neutrinos (identical for all four species in \textsc{CoCoNuT}). The bottom right panel shows the central density as a function of post-bounce time.
    \label{fig:eos_differences}}
\end{figure*}

The shorter advection time scale results from a smaller average shock radius for the CMF case. The radius $r_\mathrm{sh}$ of the stalled accretion shock is set by the PNS mass and radius $M$ and $R$, the mass accretion 
rate $\dot{M}$, and the electron flavour neutrino luminosity $L_\nu$ and energy 
$E_\nu$ as $r_\mathrm{sh}\propto 
(L_{\nu_{e}} E_\nu^2)^{4/9} R^{16/9}
\dot{M}^{-2/3} M^{-1/3}$  \citep{janka_12b,mueller_15}.
Of these, the mass and mass accretion rate are primarily set by the progenitor and not influenced appreciably by the EoS prior to the onset of explosion, except for minor differences due to the impact of the EoS on the collapse time scale. The key driver of the different evolution of the CMF models are the luminosities and mean energies, which are  appreciably smaller than for SFHo and SFHx, overcompensating for the slightly bigger PNS radius and gain radius.  The larger PNS radius for the CMF models result in weaker and less energetic neutrino emission.  

\begin{figure*}
\includegraphics[width=\textwidth]{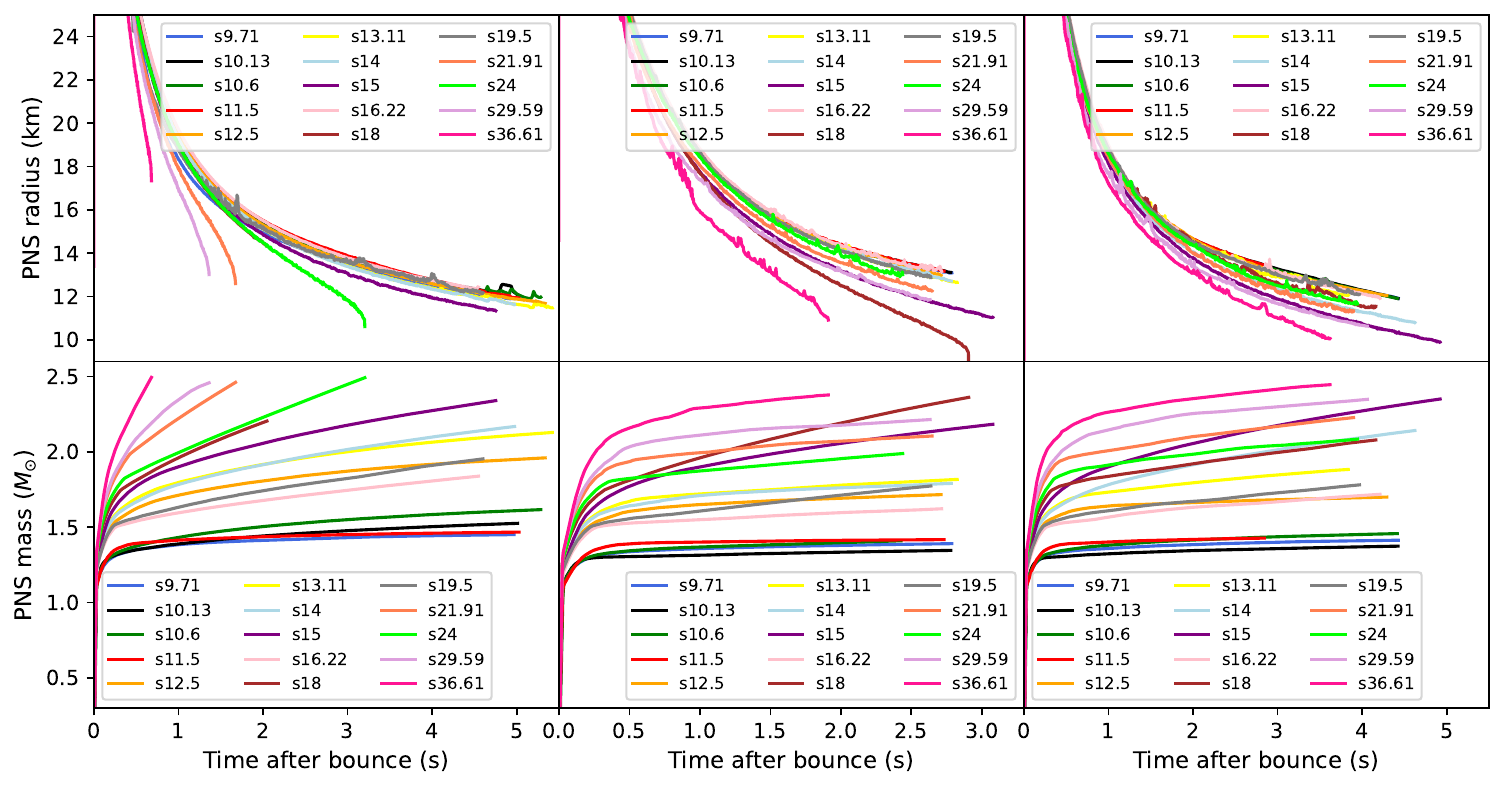}
\caption{The evolution of the PNS radius (top row), and the PNS mass (bottom row), for all models. From left to right are the CMF, SFHo and SFHx models. The highest mass CMF models form a black hole before the simulation end time. The s36.61 and s18 SFHo models also form a black hole before the simulation end time. }
\label{fig:pns_mass_radius}
\end{figure*}

The differences in PNS radii also shown in Figure \ref{fig:pns_mass_radius}, and neutrino luminosities are due to a complex combination of causal factors. While \citet{yasin_20} were able to pin down the responsible nuclear physics parameters for different PNS contraction  by directly varying these in EoS codes, this is not possible for our comparison between the CMF EoS vis \`a vis the SFHo and SFHx EoS. However, profiles of the thermodynamics quantities provide some insights into the mechanisms. Profiles for the CMF and SFHx EoS at a post-bounce time of $0.2\,\mathrm{s}$ are shown for model s12.5
in Figure~\ref{fig:eos_differences}, along with the trajectory of the central density. The CMF EoS maintains a significantly lower central density throughout the evolution. This is a consequence of the stiffness (high adiabatic index $\Gamma$) of the CMF EoS above saturation density and below the phase transition region \citep{jakobus_22}. The slower contraction of the core at supranuclear densities also translates into a slower contraction of the PNS mantle \citep{mueller_12}. Another factor adds to the slower contraction of the PNS. The adiabatic index $\Gamma$ at subnuclear densities of $10^{13}\texttt{-}10^{14}\,\mathrm{g}\,\mathrm{cm}^{-3}$ is also different between the CMF and SFHx model. This affects the structure of the PNS convection zone at these densities. Differences in the electron-flavour luminosities build up mostly in the region around 
$10^{13}\,\mathrm{g}\,\mathrm{cm}^{-3}$ at the outer edge of the convection zone.
In addition, both the core and the mantle (below nuclear density) remain cooler for the CMF model. The steeper temperature gradient for the SFHx EoS somewhat increases the heavy-flavour neutrino flux in this region. Heating due to inelastic scattering on nucleons at lower densities may also indirectly contribute to the increased electron-flavour luminosity in the SFHx model as well.

Differences in the neutrino opacities may also contribute, but we suspect that the implemented neutrino interaction rates in \textsc{CoCoNuT-FMT}, treat the opacities with reasonable consistency with the EoS in the relevant subnuclear regime. In our simulations, correlation effects are taken into account based on the virial expansion fit of \citet{horowitz_17} in terms of density, temperature and electron fraction. The fit is appropriate in the sub-nuclear regime where differences in the emerging neutrino luminosities arise.

While the larger PNS radius and gain radius for the CMF EoS also decrease the binding energy of the gain region, less heating compensates for this, and the heating time scale remains unaffected. 

Due to very long physical simulation times, we can also study some trends in explosion energies. We first note that several of our exploding high-mass progenitors reach energies above $10^{51}\,\mathrm{erg}$, despite less favourable conditions for the growth of the explosion energy in 2D \citep{mueller_15b}. Our models thus further support findings of a correlation of progenitor mass with explosion energy \citep{mueller_19,burrows_24}.
Model s36.61 even reaches $3.37\times 10^{51}\,\mathrm{erg}$, further corroborating
the notion that some neutrino-driven explosions may reach unusually high energies,
as already found in recent 3D models \citep{chan_20,burrows_24}.
However, such explosions of high-mass progenitors may, like s36.61, later form black holes by fallback \citep{chan_20,burrows_24}, which may result in a considerable reduction of the explosion energy. 

For a given progenitor, explosion energies often differ by a factor of $2\texttt{-}3$ even between the SFHo and SFHx models. We also see significant differences in explosion energies for the s18 SFHo models with different perturbations. However, even when explosion energies are computed at the same time of $1.91\,\mathrm{s}$ there is no clear EoS dependence. 
The s18 SHFo models show a larger variation in explosion energy, but we do not find any correlation with the shock revival time.
Detailed inspection of the simulation data, including the models with different perturbations, reveal considerable stochasticity in the evolution of the explosion energy, which is well known for 2D models \citep{mueller_15b}.
Furthermore, explosion energies have yet to reach their asymptotic values for many progenitors, though they appear to have saturated in some of the more massive ones. The time of shock revival may be more reflective of EoS differences. The significant delay of the explosion of 
model s9.71 for the CMF EoS by $80\texttt{-}90\,\mathrm{ms}$ compared to
the fast explosions for SFHo and SFHx is noteworthy. However, there is no clear systematic difference in the time of shock revival between the SFHo and SFHx models, which generally explode at a similar time within a few ten $\mathrm{ms}$, with the exception of a substantial delay for model s24 with the SFHx EoS.

\section{Discussion}
\label{sec:discussion}
Our results clearly demonstrate that even with tighter constraints on the maximum neutron star mass, neutron star radii, and physical properties of nuclear matter, the EoS still has a large impact on the fate of CCSNe, and hence the explosion dynamics and remnant properties.
Even for the very similar SFHo and SFHx EoS, the simulations disagree about the outcome of neutron star vs.\ black hole formation in 2 out of 15 cases.
If we denote the fraction of exploding models for each EoS and progenitor $i$ with $P_{\mathrm{SFHo},i}$ and $P_{\mathrm{SFHx},i}$, we can compute the agreement between models as
$ 1- N^{-1}\sum_{i} |P_{\mathrm{SFHo},i}-P_{\mathrm{SFHx},i}|$, 
where $N$ is the number of progenitors. Based on this metric, the outcomes with SFHo and SFHx agree with 13.3\%.
But the CMF models suggest that the EoS can still make a difference between almost no explosions (CMF) and successful explosions in almost all cases (SFHo and SFHx).

This implies that it is not yet possible to quantitatively predict the progenitor mass range(s) for successful explosions and for neutron star and black hole formation from first-principle simulations. Many qualitative results of
recent impressive parameter studies of the explodability and explosion parameters of massive stars in 2D and 3D may still be robust as they reflect a self-regulation behaviour of neutrino-driven explosions, such as correlations between progenitor core mass, explosion energy and kick velocity in the event of successful explosions
\citep{mueller_19,burrows_24}. But accurate quantitative predictions require the reduction of uncertainties in the EoS.

The EoS will obviously critically influence the distribution of neutron star and black hole masses, which is revealed in increasing details by electromagnetic \citep{ozel_16, you_25}
and gravitational-wave observations \citep{gwtc-4}. 
Importantly, the impact of the EoS is not only mediated by changing mass ranges for successful explosions. Our simulations differ from other recent sets  
\citep{burrows_25} 
in that we also see quiet black-hole formation without prior shock revival in high-mass progenitors for the CMF models and for s18 with the SFHo EoS, whereas \citet{burrows_25} only found this behaviour (their Channel~4) for low masses. This implies that the scenario of successful fallback explosions for high masses \citep{chan_20,burrows_25} is sensitive to the EoS, and with it the production channel of relatively light black holes like in the gravitational-wave binary GW190814 \citep{antoniadis_22}.
 
It is also noteworthy that the non-exploding models show sustained standing accretion shock instability (SASI) activity over long durations.
This is a consequence of the small shock radii and the short advection time scales \citep{foglizzo_06, mueller_12}.
Such long SASI activity will imprint distinct features in the gravitational-wave \citep{kuroda_16}
and neutrino signal \citep{tamborra_13, mueller_19b},
and the long duration of the signal may help to reach sufficiently high signal-to-noise ratios for detection. It remains to be seen whether multi-messenger observations of an electromagnetically quiet
Galactic supernova can constrain down the EoS using these signal features. 

It is important to note that our study shows a strong \emph{sensitivity} of explosion outcomes to the EoS, but that the almost complete lack of explosions for the CMF EoS and the very robust explosions for SFHo and SFHx are \emph{not} firm predictions due to approximations and assumptions made in our models. As our simulations are 2D, the results are not impacted by 3D effects, for example spiral SASI modes. We also neglect the impact of rotation and magnetic fields on the different EoS. Detailed predictions of outcomes will depend on 3D effects, more and more detailed micro- and macrophysics, such as neutrino flavour conversion \citep{ehring_23},
the 3D progenitor structure \citep{couch_15,mueller_17},
magnetic fields \citep{mueller_20b}
and on numerical accuracy. A wider range of successful explosions with the CMF EoS and a wider range of black hole formation cases with the SFHo and SFHx remains plausible.
\\

\section{Conclusions}
\label{sec:conclusions}
The sensitivity of explosion outcomes reinforces the need to more systematically quantify sensitivities and uncertainties of CCSN outcomes to the details of the multi-physics problem of the explosion mechanism. This is required both for forward predictions and for controlling for confounding factors when inferring CCSN or stellar evolution physics from the population of explosion and compact remnant properties. In the future, it will be important to scan the range of allowed EoS more thoroughly and to also use neutrino opacities fully consistent with the EoS, which the current study cannot yet accomplish. The nuclear EoS thus remains a major factor in the CCSN explosion problem.

\section{Data availability} The data from these simulations will be made available upon reasonable request to the authors. 

\section{Acknowledgments} The authors are supported by the Australian Research Council's (ARC) Centre of Excellence for Gravitational Wave Discovery (OzGrav) through project number CE230100016. BM acknowledges support from the ARC through Discovery Project DP240101786. JP acknowledges support from the ARC through LIEF Project LE260100008. The authors acknowledge computer time allocations from Astronomy Australia Limited's ASTAC scheme, the National Computational Merit Allocation Scheme (NCMAS), and from an Australasian Leadership Computing Grant. Some of this work was performed on the Gadi supercomputer with the assistance of resources and services from the National Computational Infrastructure (NCI), which is supported by the Australian Government, and through support by an Australasian Leadership Computing Grant.  Some of this work was performed on the OzSTAR national facility at Swinburne University of Technology. The OzSTAR program receives funding in part from the Astronomy National Collaborative Research Infrastructure Strategy (NCRIS) allocation provided by the Australian Government, and from the Victorian Higher Education State Investment Fund (VHESIF) provided by the Victorian Government.



\bibliographystyle{apsrev}

\bibliography{example} 

@ARTICLE{Anderson_21,
       author = {{Eggenberger Andersen}, Oliver and {Zha}, Shuai and {da Silva Schneider}, Andr{\'e} and {Betranhandy}, Aurore and {Couch}, Sean M. and {O'Connor}, Evan P.},
        title = "{Equation-of-state Dependence of Gravitational Waves in Core-collapse Supernovae}",
      journal = {\apj},
         year = 2021,
        month = dec,
       volume = {923},
       number = {2},
          eid = {201},
        pages = {201},
          doi = {10.3847/1538-4357/ac294c},
archivePrefix = {arXiv},
       eprint = {2106.09734},
 primaryClass = {astro-ph.HE},
       adsurl = {https://ui.adsabs.harvard.edu/abs/2021ApJ...923..201E}
}

@ARTICLE{Anderson_25,
       author = {{Eggenberger Andersen}, Oliver and {O'Connor}, Evan and {Andresen}, Haakon and {da Silva Schneider}, Andr{\'e} and {Couch}, Sean M.},
        title = "{Black Hole Supernovae, Their Equation of State Dependence, and Ejecta Composition}",
      journal = {\apj},
         year = 2025,
        month = feb,
       volume = {980},
       number = {1},
          eid = {53},
        pages = {53},
          doi = {10.3847/1538-4357/ada899},
archivePrefix = {arXiv},
       eprint = {2411.11969},
 primaryClass = {astro-ph.HE},
       adsurl = {https://ui.adsabs.harvard.edu/abs/2025ApJ...980...53E}
}

@ARTICLE{antoniadis_22,
       author = {{Antoniadis}, John and {Aguilera-Dena}, David R. and {Vigna-G{\'o}mez}, Alejandro and {Kramer}, Michael and {Langer}, Norbert and {M{\"u}ller}, Bernhard and {Tauris}, Thomas M. and {Wang}, Chen and {Xu}, Xiao-Tian},
        title = "{Explodability fluctuations of massive stellar cores enable asymmetric compact object mergers such as GW190814}",
      journal = {\aap},
         year = 2022,
        month = jan,
       volume = {657},
          eid = {L6},
        pages = {L6},
          doi = {10.1051/0004-6361/202142322},
archivePrefix = {arXiv},
       eprint = {2110.01393},
 primaryClass = {astro-ph.HE},
       adsurl = {https://ui.adsabs.harvard.edu/abs/2022A&A...657L...6A}
}

@ARTICLE{bizouard_21,
       author = {{Bizouard}, Marie-Anne and {Maturana-Russel}, Patricio and {Torres-Forn{\'e}}, Alejandro and {Obergaulinger}, Martin and {Cerd{\'a}-Dur{\'a}n}, Pablo and {Christensen}, Nelson and {Font}, Jos{\'e} A. and {Meyer}, Renate},
        title = "{Inference of protoneutron star properties from gravitational-wave data in core-collapse supernovae}",
      journal = {\prd},
         year = 2021,
        month = mar,
       volume = {103},
       number = {6},
          eid = {063006},
        pages = {063006},
          doi = {10.1103/PhysRevD.103.063006},
archivePrefix = {arXiv},
       eprint = {2012.00846},
 primaryClass = {gr-qc},
       adsurl = {https://ui.adsabs.harvard.edu/abs/2021PhRvD.103f3006B}
}

@ARTICLE{buras_06,
       author = {{Buras}, R. and {Rampp}, M. and {Janka}, H. -Th. and {Kifonidis}, K.},
        title = "{Two-dimensional hydrodynamic core-collapse supernova simulations with spectral neutrino transport. I. Numerical method and results for a 15 M{\ensuremath{\odot}} star}",
      journal = {\aap},
         year = 2006,
        month = mar,
       volume = {447},
       number = {3},
        pages = {1049-1092},
          doi = {10.1051/0004-6361:20053783},
archivePrefix = {arXiv},
       eprint = {astro-ph/0507135},
 primaryClass = {astro-ph},
       adsurl = {https://ui.adsabs.harvard.edu/abs/2006A&A...447.1049B}
}

@ARTICLE{burrows_07,
       author = {{Burrows}, A. and {Dessart}, L. and {Livne}, E. and {Ott}, C.~D. and {Murphy}, J.},
        title = "{Simulations of Magnetically Driven Supernova and Hypernova Explosions in the Context of Rapid Rotation}",
      journal = {\apj},
         year = 2007,
        month = jul,
       volume = {664},
       number = {1},
        pages = {416-434},
          doi = {10.1086/519161},
archivePrefix = {arXiv},
       eprint = {astro-ph/0702539},
 primaryClass = {astro-ph},
       adsurl = {https://ui.adsabs.harvard.edu/abs/2007ApJ...664..416B}
}

@ARTICLE{burrows_20,
       author = {{Burrows}, Adam and {Radice}, David and {Vartanyan}, David and {Nagakura}, Hiroki and {Skinner}, M. Aaron and {Dolence}, Joshua C.},
        title = "{The overarching framework of core-collapse supernova explosions as revealed by 3D FORNAX simulations}",
      journal = {\mnras},
         year = 2020,
        month = jan,
       volume = {491},
       number = {2},
        pages = {2715-2735},
          doi = {10.1093/mnras/stz3223},
archivePrefix = {arXiv},
       eprint = {1909.04152},
 primaryClass = {astro-ph.HE},
       adsurl = {https://ui.adsabs.harvard.edu/abs/2020MNRAS.491.2715B}
}

@ARTICLE{burrows_24,
       author = {{Burrows}, Adam and {Wang}, Tianshu and {Vartanyan}, David},
        title = "{Physical Correlations and Predictions Emerging from Modern Core-collapse Supernova Theory}",
      journal = {\apjl},
         year = 2024,
        month = mar,
       volume = {964},
       number = {1},
          eid = {L16},
        pages = {L16},
          doi = {10.3847/2041-8213/ad319e},
archivePrefix = {arXiv},
       eprint = {2401.06840},
 primaryClass = {astro-ph.HE},
       adsurl = {https://ui.adsabs.harvard.edu/abs/2024ApJ...964L..16B}
}

@ARTICLE{burrows_25,
       author = {{Burrows}, Adam and {Wang}, Tianshu and {Vartanyan}, David},
        title = "{Channels of Stellar-mass Black Hole Formation}",
      journal = {\apj},
         year = 2025,
        month = jul,
       volume = {987},
       number = {2},
          eid = {164},
        pages = {164},
          doi = {10.3847/1538-4357/addd04},
archivePrefix = {arXiv},
       eprint = {2412.07831},
 primaryClass = {astro-ph.SR},
       adsurl = {https://ui.adsabs.harvard.edu/abs/2025ApJ...987..164B}
}

@ARTICLE{chan_20,
       author = {{Chan}, Conrad and {M{\"u}ller}, Bernhard and {Heger}, Alexander},
        title = "{The impact of fallback on the compact remnants and chemical yields of core-collapse supernovae}",
      journal = {\mnras},
         year = 2020,
        month = jul,
       volume = {495},
       number = {4},
        pages = {3751-3762},
          doi = {10.1093/mnras/staa1431},
archivePrefix = {arXiv},
       eprint = {2003.04320},
 primaryClass = {astro-ph.SR},
       adsurl = {https://ui.adsabs.harvard.edu/abs/2020MNRAS.495.3751C}
}

@ARTICLE{couch_15,
       author = {{Couch}, Sean M. and {Chatzopoulos}, Emmanouil and {Arnett}, W. David and {Timmes}, F.~X.},
        title = "{The Three-dimensional Evolution to Core Collapse of a Massive Star}",
      journal = {\apjl},
         year = 2015,
        month = jul,
       volume = {808},
       number = {1},
          eid = {L21},
        pages = {L21},
          doi = {10.1088/2041-8205/808/1/L21},
archivePrefix = {arXiv},
       eprint = {1503.02199},
 primaryClass = {astro-ph.HE},
       adsurl = {https://ui.adsabs.harvard.edu/abs/2015ApJ...808L..21C}
}

@ARTICLE{couch_20,
       author = {{Couch}, Sean M. and {Warren}, MacKenzie L. and {O'Connor}, Evan P.},
        title = "{Simulating Turbulence-aided Neutrino-driven Core-collapse Supernova Explosions in One Dimension}",
      journal = {\apj},
         year = 2020,
        month = feb,
       volume = {890},
       number = {2},
          eid = {127},
        pages = {127},
          doi = {10.3847/1538-4357/ab609e},
archivePrefix = {arXiv},
       eprint = {1902.01340},
 primaryClass = {astro-ph.HE},
       adsurl = {https://ui.adsabs.harvard.edu/abs/2020ApJ...890..127C}
}

@ARTICLE{drischler_17,
       author = {{Drischler}, C. and {Hebeler}, K. and {Schwenk}, A.},
        title = "{Chiral interactions up to next-to-next-to-next-to-leading order and nuclear saturation}",
      journal = {arXiv e-prints},
         year = 2017,
        month = oct,
          eid = {arXiv:1710.08220},
        pages = {arXiv:1710.08220},
          doi = {10.48550/arXiv.1710.08220},
archivePrefix = {arXiv},
       eprint = {1710.08220},
 primaryClass = {nucl-th},
       adsurl = {https://ui.adsabs.harvard.edu/abs/2017arXiv171008220D}
}

@ARTICLE{ehring_23,
       author = {{Ehring}, Jakob and {Abbar}, Sajad and {Janka}, Hans-Thomas and {Raffelt}, Georg and {Tamborra}, Irene},
        title = "{Fast Neutrino Flavor Conversions Can Help and Hinder Neutrino-Driven Explosions}",
      journal = {\prl},
         year = 2023,
        month = aug,
       volume = {131},
       number = {6},
          eid = {061401},
        pages = {061401},
          doi = {10.1103/PhysRevLett.131.061401},
archivePrefix = {arXiv},
       eprint = {2305.11207},
 primaryClass = {astro-ph.HE},
       adsurl = {https://ui.adsabs.harvard.edu/abs/2023PhRvL.131f1401E}
}

@ARTICLE{ertl_16,
       author = {{Ertl}, T. and {Janka}, H.-Th. and {Woosley}, S.~E. and {Sukhbold}, T. and {Ugliano}, M.},
        title = "{A Two-parameter Criterion for Classifying the Explodability of Massive Stars by the Neutrino-driven Mechanism}",
      journal = {\apj},
         year = 2016,
        month = feb,
       volume = {818},
       number = {2},
          eid = {124},
        pages = {124},
          doi = {10.3847/0004-637X/818/2/124},
archivePrefix = {arXiv},
       eprint = {1503.07522},
 primaryClass = {astro-ph.SR},
       adsurl = {https://ui.adsabs.harvard.edu/abs/2016ApJ...818..124E}
}

@ARTICLE{fischer_18,
       author = {{Fischer}, Tobias and {Bastian}, Niels-Uwe F. and {Wu}, Meng-Ru and {Baklanov}, Petr and {Sorokina}, Elena and {Blinnikov}, Sergei and {Typel}, Stefan and {Kl{\"a}hn}, Thomas and {Blaschke}, David B.},
        title = "{Quark deconfinement as a supernova explosion engine for massive blue supergiant stars}",
      journal = {Nature Astronomy},
         year = 2018,
        month = oct,
       volume = {2},
        pages = {980-986},
          doi = {10.1038/s41550-018-0583-0},
archivePrefix = {arXiv},
       eprint = {1712.08788},
 primaryClass = {astro-ph.HE},
       adsurl = {https://ui.adsabs.harvard.edu/abs/2018NatAs...2..980F}
}

@ARTICLE{foglizzo_06,
       author = {{Foglizzo}, T. and {Scheck}, L. and {Janka}, H. -Th.},
        title = "{Neutrino-driven Convection versus Advection in Core-Collapse Supernovae}",
      journal = {\apj},
         year = 2006,
        month = dec,
       volume = {652},
       number = {2},
        pages = {1436-1450},
          doi = {10.1086/508443},
archivePrefix = {arXiv},
       eprint = {astro-ph/0507636},
 primaryClass = {astro-ph},
       adsurl = {https://ui.adsabs.harvard.edu/abs/2006ApJ...652.1436F}
}

@ARTICLE{ghosh_22,
       author = {{Ghosh}, Somdutta and {Wolfe}, Noah and {Fr{\"o}hlich}, Carla},
        title = "{PUSHing Core-collapse Supernovae to Explosions in Spherical Symmetry. V. Equation of State Dependency of Explosion Properties, Nucleosynthesis Yields, and Compact Remnants}",
      journal = {\apj},
         year = 2022,
        month = apr,
       volume = {929},
       number = {1},
          eid = {43},
        pages = {43},
          doi = {10.3847/1538-4357/ac4d20},
archivePrefix = {arXiv},
       eprint = {2107.13016},
 primaryClass = {astro-ph.HE},
       adsurl = {https://ui.adsabs.harvard.edu/abs/2022ApJ...929...43G}
}

@ARTICLE{gwtc-4,
       author = {{The LIGO Scientific Collaboration} and {the Virgo Collaboration} and {the KAGRA Collaboration} and {Abac}, A.~G. and {Abouelfettouh}, I. and {Acernese}, F. and {Ackley}, K. and others},
        title = "{GWTC-4.0: Updating the Gravitational-Wave Transient Catalog with Observations from the First Part of the Fourth LIGO-Virgo-KAGRA Observing Run}",
      journal = {arXiv e-prints},
         year = 2025,
        month = aug,
          eid = {arXiv:2508.18082},
        pages = {arXiv:2508.18082},
          doi = {10.48550/arXiv.2508.18082},
archivePrefix = {arXiv},
       eprint = {2508.18082},
 primaryClass = {gr-qc},
       adsurl = {https://ui.adsabs.harvard.edu/abs/2025arXiv250818082T}
}

@ARTICLE{heger_10,
       author = {{Heger}, Alexander and {Woosley}, S.~E.},
        title = "{Nucleosynthesis and Evolution of Massive Metal-free Stars}",
      journal = {\apj},
         year = 2010,
        month = nov,
       volume = {724},
       number = {1},
        pages = {341-373},
          doi = {10.1088/0004-637X/724/1/341},
archivePrefix = {arXiv},
       eprint = {0803.3161},
 primaryClass = {astro-ph},
       adsurl = {https://ui.adsabs.harvard.edu/abs/2010ApJ...724..341H}
}

@ARTICLE{horowitz_17,
       author = {{Horowitz}, C.~J. and {Caballero}, O.~L. and {Lin}, Zidu and {O'Connor}, Evan and {Schwenk}, A.},
        title = "{Neutrino-nucleon scattering in supernova matter from the virial expansion}",
      journal = {\prc},
         year = 2017,
        month = feb,
       volume = {95},
       number = {2},
          eid = {025801},
        pages = {025801},
          doi = {10.1103/PhysRevC.95.025801},
archivePrefix = {arXiv},
       eprint = {1611.05140},
 primaryClass = {nucl-th},
       adsurl = {https://ui.adsabs.harvard.edu/abs/2017PhRvC..95b5801H}
}

@ARTICLE{ivanov_21,
       author = {{Ivanov}, Mario and {Fern{\'a}ndez}, Rodrigo},
        title = "{Mass Ejection in Failed Supernovae: Equation of State and Neutrino Loss Dependence}",
      journal = {\apj},
         year = 2021,
        month = apr,
       volume = {911},
       number = {1},
          eid = {6},
        pages = {6},
          doi = {10.3847/1538-4357/abe59e},
archivePrefix = {arXiv},
       eprint = {2101.02712},
 primaryClass = {astro-ph.HE},
       adsurl = {https://ui.adsabs.harvard.edu/abs/2021ApJ...911....6I}
}

@INCOLLECTION{janka_17,
       author = {{Janka}, Hans-Thomas},
        title = "{Neutrino-Driven Explosions}",
    booktitle = {Handbook of Supernovae},
         year = 2017,
       editor = {{Alsabti}, Athem W. and {Murdin}, Paul},
        pages = {1095},
          doi = {10.1007/978-3-319-21846-5_109},
       adsurl = {https://ui.adsabs.harvard.edu/abs/2017hsn..book.1095J}
}

@ARTICLE{janka_01,
       author = {{Janka}, H. -Th.},
        title = "{Conditions for shock revival by neutrino heating in core-collapse supernovae}",
      journal = {\aap},
         year = 2001,
        month = mar,
       volume = {368},
        pages = {527-560},
          doi = {10.1051/0004-6361:20010012},
archivePrefix = {arXiv},
       eprint = {astro-ph/0008432},
 primaryClass = {astro-ph},
       adsurl = {https://ui.adsabs.harvard.edu/abs/2001A&A...368..527J}
}

@ARTICLE{janka_12,
       author = {{Janka}, Hans-Thomas and {Hanke}, Florian and {H{\"u}depohl}, Lorenz and {Marek}, Andreas and {M{\"u}ller}, Bernhard and {Obergaulinger}, Martin},
        title = "{Core-collapse supernovae: Reflections and directions}",
      journal = {Progress of Theoretical and Experimental Physics},
         year = 2012,
        month = dec,
       volume = {2012},
       number = {1},
          eid = {01A309},
        pages = {01A309},
          doi = {10.1093/ptep/pts067},
archivePrefix = {arXiv},
       eprint = {1211.1378},
 primaryClass = {astro-ph.SR},
       adsurl = {https://ui.adsabs.harvard.edu/abs/2012PTEP.2012aA309J}
}

@ARTICLE{janka_12b,
       author = {{Janka}, Hans-Thomas},
        title = "{Explosion Mechanisms of Core-Collapse Supernovae}",
      journal = {Annual Review of Nuclear and Particle Science},
         year = 2012,
        month = nov,
       volume = {62},
       number = {1},
        pages = {407-451},
          doi = {10.1146/annurev-nucl-102711-094901},
archivePrefix = {arXiv},
       eprint = {1206.2503},
 primaryClass = {astro-ph.SR},
       adsurl = {https://ui.adsabs.harvard.edu/abs/2012ARNPS..62..407J}
}

@ARTICLE{jakobus_22,
       author = {{Jakobus}, Pia and {M{\"u}ller}, Bernhard and {Heger}, Alexander and {Motornenko}, Anton and {Steinheimer}, Jan and {Stoecker}, Horst},
        title = "{The role of the hadron-quark phase transition in core-collapse supernovae}",
      journal = {\mnras},
         year = 2022,
        month = oct,
       volume = {516},
       number = {2},
        pages = {2554-2574},
          doi = {10.1093/mnras/stac2352},
archivePrefix = {arXiv},
       eprint = {2204.10397},
 primaryClass = {astro-ph.HE},
       adsurl = {https://ui.adsabs.harvard.edu/abs/2022MNRAS.516.2554J}
}

@ARTICLE{jakobus_23,
       author = {{Jakobus}, Pia and {M{\"u}ller}, Bernhard and {Heger}, Alexander and {Zha}, Shuai and {Powell}, Jade and {Motornenko}, Anton and {Steinheimer}, Jan and {St{\"o}cker}, Horst},
        title = "{Gravitational Waves from a Core g Mode in Supernovae as Probes of the High-Density Equation of State}",
      journal = {\prl},
         year = 2023,
        month = nov,
       volume = {131},
       number = {19},
          eid = {191201},
        pages = {191201},
          doi = {10.1103/PhysRevLett.131.191201},
archivePrefix = {arXiv},
       eprint = {2301.06515},
 primaryClass = {astro-ph.HE},
       adsurl = {https://ui.adsabs.harvard.edu/abs/2023PhRvL.131s1201J}
}

@ARTICLE{kuroda_16,
       author = {{Kuroda}, Takami and {Kotake}, Kei and {Takiwaki}, Tomoya},
        title = "{A New Gravitational-wave Signature from Standing Accretion Shock Instability in Supernovae}",
      journal = {\apjl},
         year = 2016,
        month = sep,
       volume = {829},
       number = {1},
          eid = {L14},
        pages = {L14},
          doi = {10.3847/2041-8205/829/1/L14},
archivePrefix = {arXiv},
       eprint = {1605.09215},
 primaryClass = {astro-ph.HE},
       adsurl = {https://ui.adsabs.harvard.edu/abs/2016ApJ...829L..14K}
}

@ARTICLE{lattimer_91,
       author = {{Lattimer}, James M. and {Swesty}, Douglas F.},
        title = "{A generalized equation of state for hot, dense matter}",
      journal = {\nphysa},
         year = 1991,
        month = dec,
       volume = {535},
       number = {2},
        pages = {331-376},
          doi = {10.1016/0375-9474(91)90452-C},
       adsurl = {https://ui.adsabs.harvard.edu/abs/1991NuPhA.535..331L}
}

@ARTICLE{lattimer_13,
       author = {{Lattimer}, James M. and {Lim}, Yeunhwan},
        title = "{Constraining the Symmetry Parameters of the Nuclear Interaction}",
      journal = {\apj},
         year = 2013,
        month = jul,
       volume = {771},
       number = {1},
          eid = {51},
        pages = {51},
          doi = {10.1088/0004-637X/771/1/51},
archivePrefix = {arXiv},
       eprint = {1203.4286},
 primaryClass = {nucl-th},
       adsurl = {https://ui.adsabs.harvard.edu/abs/2013ApJ...771...51L}
}

@ARTICLE{marek_09,
       author = {{Marek}, A. and {Janka}, H. -T. and {M{\"u}ller}, E.},
        title = "{Equation-of-state dependent features in shock-oscillation modulated neutrino and gravitational-wave signals from supernovae}",
      journal = {\aap},
         year = 2009,
        month = mar,
       volume = {496},
       number = {2},
        pages = {475-494},
          doi = {10.1051/0004-6361/200810883},
archivePrefix = {arXiv},
       eprint = {0808.4136},
 primaryClass = {astro-ph},
       adsurl = {https://ui.adsabs.harvard.edu/abs/2009A&A...496..475M}
}

@ARTICLE{meskhi_22,
       author = {{Meskhi}, Mikhail M. and {Wolfe}, Noah E. and {Dai}, Zhenyu and {Fr{\"o}hlich}, Carla and {Miller}, Jonah M. and {Wong}, Raymond K.~W. and {Vilalta}, Ricardo},
        title = "{A New Constraint on the Nuclear Equation of State from Statistical Distributions of Compact Remnants of Supernovae}",
      journal = {\apjl},
         year = 2022,
        month = jun,
       volume = {932},
       number = {1},
          eid = {L3},
        pages = {L3},
          doi = {10.3847/2041-8213/ac7054},
archivePrefix = {arXiv},
       eprint = {2111.01815},
 primaryClass = {astro-ph.HE},
       adsurl = {https://ui.adsabs.harvard.edu/abs/2022ApJ...932L...3M}
}

@ARTICLE{mezzacappa_20,
       author = {{Mezzacappa}, Anthony and {Endeve}, Eirik and {Messer}, O.~E. Bronson and {Bruenn}, Stephen W.},
        title = "{Physical, numerical, and computational challenges of modeling neutrino transport in core-collapse supernovae}",
      journal = {Living Reviews in Computational Astrophysics},
         year = 2020,
        month = dec,
       volume = {6},
       number = {1},
          eid = {4},
        pages = {4},
          doi = {10.1007/s41115-020-00010-8},
archivePrefix = {arXiv},
       eprint = {2010.09013},
 primaryClass = {astro-ph.HE},
       adsurl = {https://ui.adsabs.harvard.edu/abs/2020LRCA....6....4M}
}

@ARTICLE{mitra_24,
       author = {{Mitra}, A. and {Orel}, D. and {Abylkairov}, Y.~S. and {Shukirgaliyev}, B. and {Abdikamalov}, E.},
        title = "{Probing nuclear physics with supernova gravitational waves and machine learning}",
      journal = {\mnras},
         year = 2024,
        month = apr,
       volume = {529},
       number = {4},
        pages = {3582-3592},
          doi = {10.1093/mnras/stae714},
archivePrefix = {arXiv},
       eprint = {2310.15649},
 primaryClass = {astro-ph.HE},
       adsurl = {https://ui.adsabs.harvard.edu/abs/2024MNRAS.529.3582M}
}

@ARTICLE{mosta_14,
       author = {{M{\"o}sta}, Philipp and {Richers}, Sherwood and {Ott}, Christian D. and {Haas}, Roland and {Piro}, Anthony L. and {Boydstun}, Kristen and {Abdikamalov}, Ernazar and {Reisswig}, Christian and {Schnetter}, Erik},
        title = "{Magnetorotational Core-collapse Supernovae in Three Dimensions}",
      journal = {\apjl},
         year = 2014,
        month = apr,
       volume = {785},
       number = {2},
          eid = {L29},
        pages = {L29},
          doi = {10.1088/2041-8205/785/2/L29},
archivePrefix = {arXiv},
       eprint = {1403.1230},
 primaryClass = {astro-ph.HE},
       adsurl = {https://ui.adsabs.harvard.edu/abs/2014ApJ...785L..29M}
}

@ARTICLE{Motornenko_20,
       author = {{Motornenko}, Anton and {Steinheimer}, Jan and {Vovchenko}, Volodymyr and {Schramm}, Stefan and {Stoecker}, Horst},
        title = "{Equation of state for hot QCD and compact stars from a mean-field approach}",
      journal = {\prc},
         year = 2020,
        month = mar,
       volume = {101},
       number = {3},
          eid = {034904},
        pages = {034904},
          doi = {10.1103/PhysRevC.101.034904},
archivePrefix = {arXiv},
       eprint = {1905.00866},
 primaryClass = {hep-ph},
       adsurl = {https://ui.adsabs.harvard.edu/abs/2020PhRvC.101c4904M}
}

@ARTICLE{mueller_10,
       author = {{M{\"u}ller}, Bernhard and {Janka}, Hans-Thomas and {Dimmelmeier}, Harald},
        title = "{A New Multi-dimensional General Relativistic Neutrino Hydrodynamic Code for Core-collapse Supernovae. I. Method and Code Tests in Spherical Symmetry}",
      journal = {\apjs},
         year = 2010,
        month = jul,
       volume = {189},
       number = {1},
        pages = {104-133},
          doi = {10.1088/0067-0049/189/1/104},
archivePrefix = {arXiv},
       eprint = {1001.4841},
 primaryClass = {astro-ph.SR},
       adsurl = {https://ui.adsabs.harvard.edu/abs/2010ApJS..189..104M}
}

@ARTICLE{mueller_12,
       author = {{M{\"u}ller}, Bernhard and {Janka}, Hans-Thomas and {Heger}, Alexander},
        title = "{New Two-dimensional Models of Supernova Explosions by the Neutrino-heating Mechanism: Evidence for Different Instability Regimes in Collapsing Stellar Cores}",
      journal = {\apj},
         year = 2012,
        month = dec,
       volume = {761},
       number = {1},
          eid = {72},
        pages = {72},
          doi = {10.1088/0004-637X/761/1/72},
archivePrefix = {arXiv},
       eprint = {1205.7078},
 primaryClass = {astro-ph.SR},
       adsurl = {https://ui.adsabs.harvard.edu/abs/2012ApJ...761...72M}
}

@ARTICLE{mueller_15,
       author = {{M{\"u}ller}, B. and {Janka}, H. -Th.},
        title = "{Non-radial instabilities and progenitor asphericities in core-collapse supernovae}",
      journal = {\mnras},
         year = 2015,
        month = apr,
       volume = {448},
       number = {3},
        pages = {2141-2174},
          doi = {10.1093/mnras/stv101},
archivePrefix = {arXiv},
       eprint = {1409.4783},
 primaryClass = {astro-ph.SR},
       adsurl = {https://ui.adsabs.harvard.edu/abs/2015MNRAS.448.2141M}
}

@ARTICLE{mueller_15b,
       author = {{M{\"u}ller}, B.},
        title = "{The dynamics of neutrino-driven supernova explosions after shock revival in 2D and 3D}",
      journal = {\mnras},
         year = 2015,
        month = oct,
       volume = {453},
       number = {1},
        pages = {287-310},
          doi = {10.1093/mnras/stv1611},
archivePrefix = {arXiv},
       eprint = {1506.05139},
 primaryClass = {astro-ph.SR},
       adsurl = {https://ui.adsabs.harvard.edu/abs/2015MNRAS.453..287M}
}

@ARTICLE{mueller_16,
       author = {{M{\"u}ller}, Bernhard and {Heger}, Alexander and {Liptai}, David and {Cameron}, Joshua B.},
        title = "{A simple approach to the supernova progenitor-explosion connection}",
      journal = {\mnras},
         year = 2016,
        month = jul,
       volume = {460},
       number = {1},
        pages = {742-764},
          doi = {10.1093/mnras/stw1083},
archivePrefix = {arXiv},
       eprint = {1602.05956},
 primaryClass = {astro-ph.SR},
       adsurl = {https://ui.adsabs.harvard.edu/abs/2016MNRAS.460..742M}
}

@ARTICLE{mueller_17,
       author = {{M{\"u}ller}, Bernhard and {Melson}, Tobias and {Heger}, Alexander and {Janka}, Hans-Thomas},
        title = "{Supernova simulations from a 3D progenitor model - Impact of perturbations and evolution of explosion properties}",
      journal = {\mnras},
         year = 2017,
        month = nov,
       volume = {472},
       number = {1},
        pages = {491-513},
          doi = {10.1093/mnras/stx1962},
archivePrefix = {arXiv},
       eprint = {1705.00620},
 primaryClass = {astro-ph.SR},
       adsurl = {https://ui.adsabs.harvard.edu/abs/2017MNRAS.472..491M}
}

@ARTICLE{mueller_19,
       author = {{M{\"u}ller}, Bernhard and {Tauris}, Thomas M. and {Heger}, Alexander and {Banerjee}, Projjwal and {Qian}, Yong-Zhong and {Powell}, Jade and {Chan}, Conrad and {Gay}, Daniel W. and {Langer}, Norbert},
        title = "{Three-dimensional simulations of neutrino-driven core-collapse supernovae from low-mass single and binary star progenitors}",
      journal = {\mnras},
         year = 2019,
        month = apr,
       volume = {484},
       number = {3},
        pages = {3307-3324},
          doi = {10.1093/mnras/stz216},
archivePrefix = {arXiv},
       eprint = {1811.05483},
 primaryClass = {astro-ph.SR},
       adsurl = {https://ui.adsabs.harvard.edu/abs/2019MNRAS.484.3307M}
}

@ARTICLE{mueller_19b,
       author = {{M{\"u}ller}, B.},
        title = "{Neutrino Emission as Diagnostics of Core-Collapse Supernovae}",
      journal = {Annual Review of Nuclear and Particle Science},
         year = 2019,
        month = oct,
       volume = {69},
        pages = {253-278},
          doi = {10.1146/annurev-nucl-101918-023434},
archivePrefix = {arXiv},
       eprint = {1904.11067},
 primaryClass = {astro-ph.HE},
       adsurl = {https://ui.adsabs.harvard.edu/abs/2019ARNPS..69..253M}
}

@ARTICLE{mueller_20,
       author = {{M{\"u}ller}, Bernhard},
        title = "{Hydrodynamics of core-collapse supernovae and their progenitors}",
      journal = {Living Reviews in Computational Astrophysics},
         year = 2020,
        month = jun,
       volume = {6},
       number = {1},
          eid = {3},
        pages = {3},
          doi = {10.1007/s41115-020-0008-5},
archivePrefix = {arXiv},
       eprint = {2006.05083},
 primaryClass = {astro-ph.SR},
       adsurl = {https://ui.adsabs.harvard.edu/abs/2020LRCA....6....3M}
}

@ARTICLE{mueller_20b,
       author = {{M{\"u}ller}, Bernhard and {Varma}, Vishnu},
        title = "{A 3D simulation of a neutrino-driven supernova explosion aided by convection and magnetic fields}",
      journal = {\mnras},
         year = 2020,
        month = nov,
       volume = {498},
       number = {1},
        pages = {L109-L113},
          doi = {10.1093/mnrasl/slaa137},
archivePrefix = {arXiv},
       eprint = {2007.04775},
 primaryClass = {astro-ph.HE},
       adsurl = {https://ui.adsabs.harvard.edu/abs/2020MNRAS.498L.109M}
}

@ARTICLE{muller_24,
       author = {{M{\"u}ller}, Bernhard},
        title = "{Supernova Simulations}",
      journal = {arXiv e-prints},
         year = 2024,
        month = mar,
          eid = {arXiv:2403.18952},
        pages = {arXiv:2403.18952},
          doi = {10.48550/arXiv.2403.18952},
archivePrefix = {arXiv},
       eprint = {2403.18952},
 primaryClass = {astro-ph.HE},
       adsurl = {https://ui.adsabs.harvard.edu/abs/2024arXiv240318952M}
}

@ARTICLE{mueller_25,
       author = {{M{\"u}ller}, Bernhard and {Heger}, Alexander and {Powell}, Jade},
        title = "{Minimum Neutron Star Mass in Neutrino-Driven Supernova Explosions}",
      journal = {\prl},
         year = 2025,
        month = feb,
       volume = {134},
       number = {7},
          eid = {071403},
        pages = {071403},
          doi = {10.1103/PhysRevLett.134.071403},
archivePrefix = {arXiv},
       eprint = {2407.08407},
 primaryClass = {astro-ph.HE},
       adsurl = {https://ui.adsabs.harvard.edu/abs/2025PhRvL.134g1403M}
}

@ARTICLE{murphy_24,
       author = {{Murphy}, R. Daniel and {Casallas-Lagos}, Alejandro and {Mezzacappa}, Anthony and {Zanolin}, Michele and {Landfield}, Ryan E. and {Lentz}, Eric J. and {Marronetti}, Pedro and {Antelis}, Javier M. and {Moreno}, Claudia},
        title = "{Dependence of the reconstructed core-collapse supernova gravitational wave high-frequency feature on the nuclear equation of state in real interferometric data}",
      journal = {\prd},
         year = 2024,
        month = oct,
       volume = {110},
       number = {8},
          eid = {083006},
        pages = {083006},
          doi = {10.1103/PhysRevD.110.083006},
archivePrefix = {arXiv},
       eprint = {2406.01784},
 primaryClass = {astro-ph.HE},
       adsurl = {https://ui.adsabs.harvard.edu/abs/2024PhRvD.110h3006M}
}

@ARTICLE{nakamura_15,
       author = {{Nakamura}, Ko and {Takiwaki}, Tomoya and {Kuroda}, Takami and {Kotake}, Kei},
        title = "{Systematic features of axisymmetric neutrino-driven core-collapse supernova models in multiple progenitors}",
      journal = {\pasj},
         year = 2015,
        month = dec,
       volume = {67},
       number = {6},
          eid = {107},
        pages = {107},
          doi = {10.1093/pasj/psv073},
archivePrefix = {arXiv},
       eprint = {1406.2415},
 primaryClass = {astro-ph.HE},
       adsurl = {https://ui.adsabs.harvard.edu/abs/2015PASJ...67..107N}
}

@ARTICLE{ozel_16,
       author = {{{\"O}zel}, Feryal and {Freire}, Paulo},
        title = "{Masses, Radii, and the Equation of State of Neutron Stars}",
      journal = {\araa},
         year = 2016,
        month = sep,
       volume = {54},
        pages = {401-440},
          doi = {10.1146/annurev-astro-081915-023322},
archivePrefix = {arXiv},
       eprint = {1603.02698},
 primaryClass = {astro-ph.HE},
       adsurl = {https://ui.adsabs.harvard.edu/abs/2016ARA&A..54..401O}
}

@ARTICLE{pan_18,
       author = {{Pan}, Kuo-Chuan and {Liebend{\"o}rfer}, Matthias and {Couch}, Sean M. and {Thielemann}, Friedrich-Karl},
        title = "{Equation of State Dependent Dynamics and Multi-messenger Signals from Stellar-mass Black Hole Formation}",
      journal = {\apj},
         year = 2018,
        month = apr,
       volume = {857},
       number = {1},
          eid = {13},
        pages = {13},
          doi = {10.3847/1538-4357/aab71d},
archivePrefix = {arXiv},
       eprint = {1710.01690},
 primaryClass = {astro-ph.HE},
       adsurl = {https://ui.adsabs.harvard.edu/abs/2018ApJ...857...13P}
}

@ARTICLE{raithel_19,
       author = {{Raithel}, Carolyn A.},
        title = "{Constraints on the neutron star equation of state from GW170817}",
      journal = {European Physical Journal A},
         year = 2019,
        month = may,
       volume = {55},
       number = {5},
          eid = {80},
        pages = {80},
          doi = {10.1140/epja/i2019-12759-5},
archivePrefix = {arXiv},
       eprint = {1904.10002},
 primaryClass = {astro-ph.HE},
       adsurl = {https://ui.adsabs.harvard.edu/abs/2019EPJA...55...80R}
}









\label{lastpage}
\end{document}